# ASGARD: A Single-cell Guided Pipeline to Aid Repurposing of Drugs

Bing He[1], Yao Xiao[1], Haodong Liang[2], Qianhui Huang[1], Yuheng Du[1], Yijun Li[1], David Garmire[3], Duxin Sun[4], Lana X. Garmire[1] [*]

1. Department of Computational Medicine and Bioinformatics, Medical School, University of Michigan, Ann Arbor, MI, USA

2. Department of Statistics, College of Literature, Science, and the Arts, University of Michigan, Ann Arbor, MI, USA

3. Department of Electrical Engineering and Computer Science, College of Engineering, University of Michigan, Ann Arbor, MI, USA

4. Department of Pharmaceutical Sciences, College of Pharmacy, University of Michigan, Ann Arbor, MI, USA

*: Correspondence to Lana X. Garmire, Department of Computational Medicine and Bioinformatics, North Campus Research Complex, University of Michigan, 1600 Huron Parkway, Ann Arbor, MI 48105, USA

Telephone: (734) 615-0514

Email: lgarmire@med.umich.edu

# Abstract

Intercellular heterogeneity is a major obstacle to successful precision medicine. Single-cell RNA sequencing (scRNA-seq) technology has enabled in-depth analysis of intercellular heterogeneity in various diseases. However, its full potential for precision medicine has yet to be reached. Towards this, we propose a new drug recommendation system called: A Single-cell Guided Pipeline to Aid Repurposing of Drugs (ASGARD). ASGARD defines a novel drug score predicting drugs by considering all cell clusters to address the intercellular heterogeneity within each patient. We tested ASGARD on multiple diseases, including breast cancer, acute lymphoblastic leukemia, and coronavirus disease 2019 (COVID-19). On single-drug therapy, ASGARD shows significantly better average accuracy (AUC of 0.92) compared to two other bulk-cell-based drug repurposing methods (AUC of 0.80 and 0.76). It is also considerably better (AUC of 0.82) than other cell cluster level predicting methods (AUC of 0.67 and 0.55). In addition, ASGARD is also validated by the drug response prediction method TRANSACT with Triple-Negative-Breast-Cancer patient samples. Many top-ranked drugs are either approved by FDA or in clinical trials treating corresponding diseases. In silico cell-type specific drop-out experiments using triple-negative breast cancers show the importance of T cells in the tumor microenvironment in affecting drug predictions. In conclusion, ASGARD is a promising drug repurposing recommendation tool guided by single-cell RNA-seq for personalized medicine. ASGARD is free for educational use at https://github.com/lanagarmire/ASGARD.

# Introduction

Heterogeneity, or more specifically, the diverse cell populations within the diseased tissue, is the leading cause of treatment failure for many complex diseases, such as cancers [1], Alzheimer's disease [2], stroke [3], and coronavirus disease 2019 (COVID-19) [4], etc., as well as a major obstacle to successful precision medicine [5–7]. Recent significant advances in single-cell technologies, especially the single-cell RNA sequencing (scRNA-seq) technology, have enabled the analysis of intercellular heterogeneity at a very fine resolution [8,9] and helped us to have many breakthroughs in understanding disease mechanisms[10], such as breast cancer [11], liver cancer [12] and COVID-19 [13]. However, its full potential for precision medicine has not been fulfilled [14,15].

Drug repurposing (also known as drug reposition, reprofiling, or re-tasking) is a strategy to identify new drug uses outside the scope of its original medical approval or investigation [16]. So far, few drug repurposing methods have been developed to utilize the highly valuable information contained in scRNA-seq data. The pipeline by Alakwaa identifies significantly differentiated genes (DEGs) for a specific group of cells, then predicts candidate drugs for DEGs using the Connectivity Map Linked User Environment (CLUE) platform, followed by prioritizing these drugs using a comprehensive ranking score system [17]. This pipeline identified didanosine as a potential treatment for COVID-19 using scRNA-seq data [17]. Another pipeline by Guo et al. uses a simple combination of Seurat [18], a tool for scRNA-seq analysis, and CLUE to identify 281 FDA-approved drugs that can potentially be effective for treating COVID-19 [19]. In general, the above pipelines predict drugs for each cell cluster within the patient. However, in heterogeneous diseases caused by multiple types of cells, efficient drugs should be able to address multiple cell clusters[20]. Neither of these pipelines mentioned above can predict drugs for multiple cell clusters, limiting their utility in the era of precision medicine.

Here we propose A Single-cell Guided Pipeline to Aid Repurposing of Drugs (ASGARD) to overcome the issue above. ASGARD defines a novel drug score to predict drugs for multiple diseased cell clusters within each patient. The benchmarking results show that the performance of ASGARD on single drugs is more accurate and robust than other pipelines handling bulk and single-cell RNA-Seq data. We tested ASGARD on multiple cancer scRNA-Seq datasets, including patient-Derived Xenografts (PDXs) models for advanced metastatic breast cancers, Pre-T acute lymphoblastic leukemia patients, and primary tumors of Triple-Negative-Breast-Cancer (TNBC) patients.  Additionally, with the ongoing worldwide COVID-19 pandemic, we applied ASGARD to scRNA-seq data from severe COVID-19 patients and predicted potential therapies to reduce deaths of severe COVID-19 patients.

## Methods

**Single-cell RNA sequencing (scRNA-seq) data**

We obtained multiple scRNA-seq datasets from the Gene Expression Omnibus (GEO) database. ScRNA-seq data of cells from 4 Triple-Negative-Breast-Cancer (TNBC) patients and 4 healthy controls are from GSE161529. Epithelial cells from Patient-Derived Xenografts (PDXs) models of 2 patients with advanced metastatic TNBCs and adult human breast epithelial cells from 3 healthy women are from GEO with accession numbers GSE123926 [11] and GSE113197 [21], respectively. Another scRNA-seq pediatric bone marrow mononuclear cells (PBMMC) dataset from 2 Pre-T acute lymphoblastic leukemia patients and three healthy controls is from GEO with accession number GSE132509 [22]. The last set of scRNA-seq datasets are single cells from the bronchoalveolar lavage fluid (BALF) of 15 severe COVID-19 patients (4 deceased and 11 cured) from GEO with accession numbers GSE145926 [13] and GSE158055 [13,23].

**Processing of scRNA-seq data**

ASGARD accepts processed scRNA-seq data from the Seurat package [18]. In this study, genes identified in fewer than three cells are removed from the dataset. We used the same criteria as in their original studies to filter cells [11,13,21]. Preprocessing steps remove the following cells from the dataset: (1) epithelial cells from breast cancer PDXs and healthy breast tissues with fewer than 200 unique genes, (2) PBMC cells from leukemia patients and healthy controls with fewer than 200 unique genes, and (3) BALF cells from COVID-19 patients with fewer than 200 unique genes, more than 6000 unique genes or have a proportion of mitochondrial genes larger than 10% [13]. For consistency, cells from TNBC patients with fewer than 200 unique genes are also removed from the dataset [24]. We used cell cycle marker genes and linear transformation to scale the expression of each gene and remove the effects of the cell cycle on gene expression.

**Cell pairwise correspondences**

ASGARD suggests using functions from Seurat for cell pairwise correspondences. In this study, gene counts for each cell were divided by the total counts for that cell and multiplied by a scaling factor (default is set to 10000). The count matrix was then transformed by log 2(count+1) in R. To identify gene variance across cells, we first fitted a line to the relationship of log(variance) and log(mean) using local polynomial regression (loess). Then we standardized the feature values using the observed mean and expected variance (given by the fitted line). Gene variance was then calculated on the standardized values. We used the 2,000 genes with the highest standardized variance for downstream analysis. Then we identified the K-nearest neighbors (KNNs) between disease and normal cells based on the L2-normalized canonical correlation vectors (CCV). Finally, we built up the pairwise cell correspondences by identifying mutual nearest neighbors [18].

**Cell clustering and annotation**

We applied principal component analysis (PCA) from Seurat on the scaled data to perform the linear dimensional reduction. Then we used a graph-based clustering approach[18]. In this approach, we first constructed a KNN graph based on the euclidean distance in PCA space and refined the edge weights between any two cell pairs using Jaccard similarity. Then we applied the Louvain algorithm of modularity optimization to iteratively group cell pairs together. We further ran non-linear dimensional reduction (UMAP) to place similar cells within the graph-based clusters determined above together in low-dimensional space. To annotate clusters of cells, we ran an automatic annotation of single cells based on similarity to the referenced single-cell panel using the SingleR package [25]. We used the dominant cell type (>50% cells) as the cell type of the cluster.

**Drug repurposing**

ASGARD supports importing differentially expressed genes calculated from multiple external methods, including Limma [26], Seurat (Wilcoxon Rank Sum test) [18], DESeq2 [27], and edgeR [28]. The differentially expressed gene list in disease is transformed into a gene rank list. ASGARD uses 21,304 drugs/compounds with response gene expression profiles in 98 cell lines from the LINCS L1000 project [29]. A differential gene expression list in response to drug treatment is also transformed into a gene rank list. ASGARD further identifies potential candidate drugs that yield reversed gene expression patterns from those of diseased vs. normal cells, using the DrInsight package [30] (version: 0.1.1). Specifically, it identifies consistently differentially expressed genes, which are up-regulated in cells from diseased tissue but down-regulated in cells with drug treatment, or down-regulated in cells from diseased tissue but up-regulated in cells with drug treatment. It then calculates the outlier-sum (OS) statistic [31], representing the effect of reversed differential gene pattern by the drug treatment. The Kolmogorov–Smirnov test (K-S test) is then applied to the OS statistic, to show the significance level of one drug treatment relative to the background of all other drugs in the dataset. The reference drug dataset contains gene rank lists of 591,697 drug/compound treatments from the LINCS L1000 data, as mentioned above. The Benjamini-Hochberg procedure is used to adjust P-values from the K-S test to control False Discovery Rate (FDR) of multiple hypothesis testing[32].

**Drug evaluation**

ASGARD defines a novel drug score at the individual patient level (Formula 1), which calculates the drug efficacy across all single-cell clusters in a given patient's scRNA-Seq data. The drug score estimates drug efficacy using the cell type proportion, the significance of reversed differential gene expression pattern (FDR), and the ratio of reversed significantly deregulated genes over disease-related (or selected) single-cell clusters. The drug score is estimated by the following formula:

$$Drug\ score = \sum_{k=1}^{n} (\frac{Num(Cell)_k}{Num(Total.Cell)} * (-\log_{10}^{FDR_k}) * \frac{Num(ReversedGene)_k}{Num(DiseasedGene)_k}) \quad (1)$$

In this formula, $k$ is a particular single-cell cluster, $n$ represents all disease-related (or selected) single-cell clusters, $\frac{Num(Cell)_k}{Num(Total.Cell)}$ represents cellular proportion of the cluster $k$ in all diseased cells, $-\log_{10}^{FDR_k}$ represents the significance of reversed differential gene pattern in the cluster $k$ by drug treatment, and $\frac{Num(ReversedGene)_k}{Num(DiseasedGene)_k}$ represents the ratio of reversed disease-related genes by drug treatment. Specifically, $Num(Total.Cell)$ is the total number of cells in the sample and $Num(Cell)_k$ is the number of cells in the cluster $k$. $FDR_k$ is the drug's FDR-adjusted p-value (significance of reversed differential gene pattern) for cluster $k$. $Num(DiseasedGenes)_k$ is the number of significantly deregulated genes in a cluster $k$, while $Num(ReversedGenes)_k$ is the number of significantly deregulated genes in a cluster $k$ that can be reversed by the drug. To allow a comparison of drug efficacy across patients, ASGARD also provides a standardized drug score, which has a scale of 0 to 1 (Formula 2).

$$Standardized\ Drug\ Score = 1 - \frac{Rank(Drug)}{Total\ Num(Drug)} \quad (2)$$

Besides the drug score, ASGARD further provides Fisher's combined P-value [33] over the original P value of every cluster. The combined p-value is calculated as the right-tail probability $P_{x^2(2n)}(T > t)$, where $t = -2 \sum_{i=1}^{n} \log_{10}^{P_i}$. The BH FDR is used to adjust Fisher's combined P-value. The adjusted Fisher's combined P-value (FDR) is independent of the drug score. The FDR and drug score can be used together or independently for drug selection. By default, ASGARD uses the drug score for drug selection. Drugs with a higher value of drug score are supposed to have better therapeutic effects than those with a lower value.

**Benchmarking ASGARD**

We use the receiver operating characteristic curves (ROCs) and the areas under the ROC curves (AUCs) to compare the performance of ASGARD with those of the other two pipelines, as well as bulk methods. Since these pipelines/ methods report both drugs and compounds, we let ASGARD report both drugs and compounds in the comparisons with other pipelines/methods. ROCs and AUCs are calculated for each pipeline using the pROC package [34]. In ROC and AUC estimation, we regarded FDA-approved drugs and compounds used in advanced clinical trials or have been proven effective in animal models as positive cases (Supplementary Table 1), and all other drugs as negative cases. To identify drugs and compounds used in advanced clinical trials or have been proven effective in animal models, we used three databases that are ClinicalTrials.gov, PubMed, and PubChem, and all the drugs and compounds we found are listed in Supplementary Table 1.

We determined the need to assess the robustness of the three pipelines on different (1) sizes, (2) similarities, and (3) unbalances of single-cell populations. For (1), the Bootstrapping method in R [35] generated simulation data of different sizes by randomly drawing the same number of disease and normal cells from GSE123926 and GSE113197. For (2), additional simulation data are generated by adjusting the differential gene expression levels from 20% to 90% of the original differential levels of the single-cell cluster, based on GSE123926 and GSE113197. For (3), simulation data is generated by randomly drawing 5000 cells with diseased cell proportions ranging from 20% to 90%, thereby yielding unbalanced populations.

**Drug score analysis**

To examine the impact of each cell type on a drug score, we conducted in silico drop-one-out experiments, excluding one cell type from the scRNA-Seq data at a time. The difference between the new drug score and the original drug score is then calculated to reflect the contribution of each cell type to the drug prediction.

To validate ASGARD, we compared it with another drug response prediction method TRANSACT [36], on the TNBC dataset (GSE161529). Since TRANSACT is a method working on bulk gene expression data, we took the mean expression value of each gene across all cells as the pseudo-bulk expression value of that gene. To fit TRANSACT with the dataset we used, we changed two parameters, number_pc['target'] and n_pv, to 3 and maintained all other parameters at the same value as the authors' original report[36].

**Code availability**

ASGARD is available as an R package in Github (https://github.com/lanagarmire/ASGARD) under the PolyForm Noncommercial License. The true positive datasets, results, and scripts for the performance comparison and validation are available at: https://github.com/lanagarmire/Single-cell-drug-repositioning/tree/master/ROC

# Results

**Summary of A Single-cell Guided pipeline to Aid Repurposing of Drugs (ASGARD)**

Using scRNA-seq data, ASGARD repurposes drugs for disease by fully accounting for the cellular heterogeneity of patients (Figure 1, Formula 1 in **Methods**). In ASGARD, every cell cluster in the diseased sample is paired to that in the normal (or control) sample, according to "anchor" genes that are consistently expressed between diseased and normal cells. It then imports differentially expressed (DE) genes (P-value < 0.05) between the paired diseased and normal clusters in the scRNA-seq data, as determined by an DE detection method at the user's choice. These individual clusters can be optionally annotated to specific cell types. To identify drugs for each single cluster (cell type), then ASGARD uses these consistently differentially expressed genes as inputs to identify drugs that can significantly (single-cluster FDR < 0.05) reverse their expression levels in the L1000 drug response dataset[29]. To identify drugs for multiple clusters, ASGARD defines a novel drug score (Formula 1 in **Methods**) to evaluate the drug efficacy across multiple cell clusters selected by the user. The drug score estimates drug efficacy by taking into account the cell type proportion, the significance of reversing the differential gene expression pattern (single-cluster FDR) by the drug treatment in each selected cell cluster, and the ratio of significantly deregulated genes (adjust P-value < 0.05) that the drug treatment can reverse in each selected cell cluster. Finally, ASGARD uses the drug score to rank and choose drugs for the disease.

We evaluated the power of the drug score by comparing ASGARD with traditional bulk-cell-based repurposing methods and single-cell-based repurposing methods using multiple independent scRNA-seq datasets, including PDX models from advanced metastatic TNBC[11], an acute lymphoblastic leukemia dataset[22], and coronavirus disease 2019 (COVID-19) datasets [13,23] (see **Methods**).

**Comparing ASGARD to bulk-cell based repurposing methods**

Before comparing ASGARD to bulk-cell-based repurposing methods, we first evaluated several external differential expression (DE) methods, on three datasets from three diseases: advanced metastatic breast cancer [11,21], acute lymphoblastic leukemia[22], and coronavirus disease 2019[13,23] (see **Methods**). We selected Limma [26], Seurat [18], DESeq2 [27], and edgeR [28] for DE methods, given that they were top-ranked methods in a benchmark study of confronting false discoveries in single-cell differential expression[37]. We conducted systematic comparison of these methods under different modes (Supplementary Figure 2). For limma, we compared three modes: empirical Bayes without trend (Bayes), empirical Bayes approach prior trend (trend), and precision weights (voom)[26]. For Seurat, we compared three different DE tests: Wilcoxon rank-sum test (Wilcox), t-test, and logistic regression (LR). For DESeq2, we compared the Wald test (Wald) and the likelihood ratio test (LRT). For edgeR, we compared the likelihood ratio test (LRT) and the quasi-likelihood F-test (QLF). We identified DE genes using the above methods for each cell cluster as the inputs of ASGARD for drug repurposing. The subsequent drug prediction accuracies by ASGARD are determined by the receiver operating characteristic curves (ROCs) and the areas under the ROC curves (AUCs), using FDA-approved drugs and candidate drugs validated in advanced clinical trials as the positive data (see Methods).

The systematic comparison is shown in **Supplementary Figure 1**, and the results from each method under the best-performing mode are shown in **Figure 2A**. Limma (Bayes) method yields the best AUC in all three datasets ranging from (0.90-0.92), significantly (P-value < 0.05, student's t-test) better than other DE methods (Figure 1A). Seurat (Wilcox test) and edgeR perform similarly overall, where edgeR has slightly higher AUC (0.83-0.86) than Seurat (0.80-8.86). DEseq2 on the other hand, tends to generate some of the lowest AUCs in comparison. Therefore, we used DE results from the limma-Bayes package for the following analysis, while keeping other DE methods as options for the inputs to ASGARD.

To compare ASGARD with those drug repurposing methods using bulk RNA-Seq samples, we summarized scRNA-seq data into pseudo-bulk RNA-Seq data. We then applied bulk methods CLUE [38] and DrInsight [30] to the pseudo-bulk RNA-Seq query data and compared their results with ASGARD's on predicting both drugs and compounds (Figure 2B). We used the same scRNA-seq data from the same three datasets above. Since CLUE and DrInsight predict both drugs and compounds, we added compounds validated in animal models to the true positive dataset for the AUC evaluation of drug/compound predictions. As a result, the AUCs obtained from ASGARD on drugs and compounds (Figure 2B) are slightly different from those on drugs only (Figure 2A). On the breast cancer dataset, ASGARD yields an overall AUC of 0.92, much better than CLUE and DrInsight, with values of 0.74 and 0.81, respectively. On precursor T cell acute lymphoblastic leukemia data, ASGARD yields an AUC of 0.95 in drug/compound repurposing for leukemia patients, while CLUE and DrInsight achieve worse average AUCs of 0.82 and 0.73, respectively. For the COVID-19 datasets, ASGARD shows an AUC of 0.88 in drug/compound repurposing, while CLUE and DrInsight have lower AUCs of 0.85 and 0.73, respectively, for the same patients (Figure 2B). In summary, by paying attention to heterogeneity at single-cell levels, ASGARD shows much better drug repurposing predictability than methods that rely on bulk samples.

**Comparing ASGARD to other single-cell-based repurposing methods**

We also compared single drug prediction using ASGARD with two other pipelines developed by Alakwaa et al. [17,19] and Guo et al.[17,19], which were reported to handle scRNA-Seq data. Note that ASGARD offers more functionalities than those two methods. Alakwaa' and Guo' pipelines can only repurpose drug/compounds for every cluster, but not on a multi-cluster level. On the other hand, ASGARD can compute both the single-cluster-level drug significance and the multi-cluster novel drug score (Formula 1 in **Methods**). The above section shows that the ASGARD multi-cluster drug score shows AUCs of 0.92, 0.95, and 0.88 for breast cancer, leukemia, and COVID-19, respectively (Figure 2B). For a fair comparison, we further tested the single-cluster-level drug prediction accuracies of these three methods (Figure 3, Supplementary Figure 2). Even at the single-cluster-level, ASGARD still shows the best AUCs on every individual cluster from breast cancer, leukemia, and COVID-19 datasets (Figure 3). On the 8 clusters of the breast cancer dataset, ASGARD yields an averaged AUC of 0.83 (0.80-0.86), significantly better (P-value<0.001, student's t-test) than Alakwaa' and Guo' pipelines, with averaged AUC values of 0.72 (0.62-0.79) and 0.56 (0.54-0.59) respectively (Figure 3A, Supplementary Figure 2). On the 4 clusters of precursor T cell acute lymphoblastic leukemia data, ASGARD has an averaged AUC of 0.81 (0.76-0.85), again significantly better (P-value <0.001, student's t-test) than Alakwaa' and Guo' pipelines, with averaged AUC values of 0.51 (0.49-0.56) and 0.52 (0.49-0.55) respectively (Figure 3B, Supplementary Figure 2). Similar trends exist in the neutrophile, NK, T cell and monocytes that have increased cell proportions in the decreased severe vs. cured severe COVID-19 patients. While ASGARD achieves average AUCs of 0.82 (0.77-0.88), Alakwaa' and Guo' methods have reduced average AUCs of 0.72 (0.63-0.80), and 0.58 (0.55-0.62) (Figure 3C, Supplementary Figure 2). These results support the conclusion that ASGARD predicts drugs more accurately than Alakwaa' and Guo' pipelines.

Additionally, given that sample size, cell population similarity, and proportion of disease cells impact significantly on differential gene analysis[39], we further performed robustness assessments of the three pipelines across different sizes of single-cell populations, different differential levels of single-cell populations, and different proportions of diseased cells using simulation data based on GSE123926 and GSE113197 dataset (see **Methods**). The AUCs of the three single-cell drug repurposing pipelines on the simulation data show that ASGARD, as well as the other two pipelines, have very robust performance across different sizes of single-cell populations (Supplementary Figure 3A), different degrees of DE between disease and normal conditions (Supplementary Figure 3B), and different proportions of disease cells among the scRNA-Seq data (Supplementary Figure 3C).

We demonstrate that ASGARD is a promising drug recommendation pipeline through computational and clinical validation. In the following sections, we further illustrate the results of ASGARD applied to breast cancer, leukemia, and COVID-19, respectively.

**Drug repurposing for breast cancers**

To evaluate the reliability of ASGARD on breast cancer patient data, we downloaded four Triple Negative Breast Cancer (TNBC) samples along with four controls from the GSE161529 dataset [40], in order to compare with the drug prediction results from the PDX models of TNBC described earlier. After the preprocessing procedure by Seurat, the TNBC samples contain an average of 5580 cells. We aligned all 8 samples, paired the cases vs. controls, and clustered them into 6 groups: B-cell, endothelial cell, epithelial cell, macrophage, T cell, and tissue stem cell (Figure 4A). Epithelial cells are the largest group covering 45.98% of cells on average, while endothelial cells are the smallest group as expected, accounting for only 1.152% of total cells (Figure 4A). ASGARD predicted 13 drugs with significant FDR p-values in at least one of the four TNBC patients (Figure 4B). For comparison, we also performed drug predictions on the 2 PDX models of TNBC patients, using the same procedures (Supplementary Figure 4C). Of great interest, four of the most significant drugs from TNBC patients overlap with those predicted by the two PDX samples. These drugs are mebendazole, crizotinib, neratinib, and vinblastine (Figure 4B). Both neratinib and vinblastine have been proven by the FDA for the treatment of breast cancer [41,42]. Mebendazole is a well-known anti-helminthic drug with wide clinical use. It has been reported to have anti-cancer properties in preclinical studies and has been in many clinical trial studies for treating various cancers, including liver, lung cancers, and glioma [43]. Crizotinib is a receptor tyrosine kinase inhibitor showing tumor-reducing effects in vitro and in vivo [44,45]. It is now in phase 2 clinical trial for treating patients with TNBC (ClinicalTrials.gov Identifier: NCT03620643).

To show quantitatively that ASGARD prediction on the TNBC samples is valuable, we next conducted two additional sets of analyses. First, we compared its results with those using TRANSACT [36], another computational method to calculate drug sensitivity. Since the TRANSACT can only predict drugs existing in the GDSC dataset [46], thus we can only compare the subset of drugs predicted by ASGARD in the GDSC dataset. As shown in Figure 4C, ASGARD and TRANSACT results are well correlated. As ASGARD's drug score increases, the drug sensitivity in TRANSACT also increases. Second, we investigate the effect of the tumor microenvironment on the drug scores. Thus we did *in silico* drop-one-out experiment, which excluded one cell type at a time. Among all cell types in the tumor microenvironment, T cell leads to the most drastic drug score changes, as well as the most variable drug score changes among different drugs (Figure 4D). Moreover, the drug score changes also differ among the four TNBC patients, showing the sensitivity of ASGARD in personalized drug prediction.

To explore the potential molecular mechanisms of the top drug candidate, we next investigated the target genes and pathways of Mebendazole across the six cell clusters (Figure 4E & 4F). Mebendazole targets many important genes and pathways in TNBC, such as signal transducer and activator of transcription 1 (STAT1) in Toll-like receptor signaling pathway, Vascular Cell Adhesion Molecule 1 (VCAM1) in NF-kappa B signaling pathway, Matrix Metallopeptidase 14 (MMP14) in TNF signaling pathway, signal transducer and activator of transcription 2 (STAT2) in NOD-like receptor signaling pathway, cyclin-dependent kinase inhibitor 1A (CDKN1A) in PI3K-Akt signaling pathway, etc. These targeted genes and pathways are essential for the proliferation, migration, and invasion of TNBC cells[47,48], and were suggested as therapeutic drug targets for TNBC in previous studies[49,50].

**Drug repurposing for precursor T cell acute lymphoblastic leukemia (Pre-T ALL)**

We further applied ASGARD to the collected scRNA-seq data from 2 Pre-T ALL patients and three normal healthy controls[22]. ASGARD identifies eight types of cells (Figure 5A), in which T cells are further clustered into four sub-populations (Figure 5B). Cluster 1 (C1) is the largest one, covering 47.29% of cells, while cluster 4 (C4) is the smallest, accounting for only 2.11% of cells (Figure 5B). The differentially expressed genes (adjusted P-value < 0.05, Pre-T ALL vs. normal) in the T cell clusters are significantly enriched in 6 pathways, including apoptosis, cell cycle, cGMP−PKG signaling, NF−kappa B signaling, p53 signaling, and T cell receptor signaling pathways (Figure 5C).

Among the predicted drugs by ASGARD, the first candidate, tretinoin, has been approved for the treatment of leukemia [51] (Figure 5D, Supplementary Table 2). Tretinoin is a vitamin A derivative. We further explored the potential molecular mechanisms of the FDA-approved top1 candidate tretinoin. Tretinoin targets many leukemia-related genes and all the significant pathways in the 4 T cell clusters, including: the regulator MDM4 in the p53 signaling pathway, cyclin D3 (CCND3) in cell cycle and p53 signaling pathways, G protein subunit alpha q (GNAQ) and phospholipase C beta 1 (PLCB1) in the cGMP−PKG signaling pathway, Fos protooncogene (FOS) and p21 (RAC1) activated kinase 2 (PAK2) in the T cell receptor signaling pathway, spectrin alpha non-erythrocytic 1 (SPTAN1) in the apoptosis pathway, and zeta chain of T cell receptor-associated protein kinase 70 (ZAP70) in apoptosis and NF−kappa B signaling pathways (Figure 5E). All these genes and pathways were previously shown to have significance in the pathogenesis of Pre-T ALL [52–54]. The drug target genes and pathways in the T cell clusters explain why ASGARD predicts tretinoin for leukemia and how tretinoin treats leukemia.

**Drug repurposing for severe patients with coronavirus disease 2019 (COVID-19)**

The immune response activated by the SARS-CoV-2 virus infection is a double-edged sword. It protects the human body from viral infection. But the deregulated immune response in severe COVID-19 patients damages the alveolar to cause respiratory failure that kills the patients [55,56]. To find drugs that may help to reduce the mortality of severe COVID-19, we collected scRNA-seq data from the bronchoalveolar lavage fluid (BALF) of 15 severe COVID-19 patients [13,23]. Among them, 11 patients were cured (cured severe), while four died (deceased severe) afterward. To identify immune cells that correlate with the death of severe patients, we compared the scRNA-seq data between deceased severe and cured severe patients. In total, there are seven types of cells, including six types of immune cells and epithelial cell types (Figure 6A), in the BALF samples collected from severe COVID-19 patients. Monocyte is the largest T cell population in both deceased and cured severe COVID-19 patients(Figure 6B). The population of neutrophil, NK cell, T cell, and monocyte increased in deceased severe COVID patients compared to the cured ones, suggesting the important role of these four types of cells in COVID-19-related death [57–60] (Figure 6B). The differentially expressed genes (adjusted P-value <0.05, deceased severe vs cured severe) in the four types of cells are significantly enriched (adjusted P-value <0.05) in 8 pathways, including chemokine signaling, coronavirus disease−COVID−19, IL−17 signaling, JAK−STAT signaling, NF−kappa B signaling, T cell receptor signaling, TNF signaling and Toll−like receptor signaling pathways (Figure 6C). Coronavirus disease−COVID−19 pathway is the most significant pathway in these cells, as expected. Chemokine signaling, NF−kappa B signaling, TNF signaling, and Toll−like receptor signaling pathways are the most widely enriched pathways in all four types of cells. T cell receptor signaling pathway is only enriched in T cells.

We identified the differential gene expression profiles of the four cell types, including neutrophil, NK cell, T cell, and monocyte, by comparing decreased severe patients to cured severe ones. Then we put the differential gene expression profiles to ASGARD to identify drug candidates using the multi-cluster drug score. Among the predicted drugs, rescinnamine (2nd) and enalapril (4th) caught our attention (Figure 6D, Supplementary Table 3). Both rescinnamine and enalapril are angiotensin-converting enzyme (ACE) inhibitors. Angiotensin-converting enzyme 2 (ACE2) mediates the SARS-CoV-2 cell entry. It's interesting to see rescinnamine and enalapril are predicted by ASGARD for treating severe COVID-19. So we further explored their target genes and pathways in the four cell types. Rescinnamine and enalapril share most of the key genes on all the significant pathways in monocyte, NK cell, neutrophil, and T cell, respectively (Figure 6E). In monocyte, rescinnamine and enalapril share 47 key target genes, including Janus Kinase 1 (JAK1), Janus Kinase 2 (JAK2), C-C Motif Chemokine Ligand 2 (CCL2), C-C Motif Chemokine Ligand 4 (CCL4), and C-C Motif Chemokine Ligand 8 (CCL8), and all the 7 significant pathways. In NK cells, rescinnamine and enalapril share 35 key target genes from 6 significant pathways, such as JAK1, Janus Kinase 3 (JAK3), CCL4, tumor necrosis factor (TNF), and Signal Transducer And Activator Of Transcription 2 (STAT2). In neutrophils, rescinnamine and enalapril share 16 key target genes, such as CCL2, CCL8, C-X-C Motif Chemokine Ligand 8 (CXCL8) and C-X-C Motif Chemokine Ligand 10 (CXCL10), and all the 5 significant pathways. In T cell, rescinnamine and enalapril share 30 key target genes, such as CCL2, CCL8, C-X-C Motif Chemokine Ligand 9 (CXCL9), JAK3, TNF, and Lymphocyte Cytosolic Protein 2 (LCP2), and all the 6 significant pathways of T cell. The shared target genes and pathways in corresponding cells were previously shown related to death from COVID-19 [57–60].

## Discussion

This study presents a Single-cell Guided pipeline to Aid Repurposing of Drugs (ASGARD) as a new generation of personalized drug recommendation system. To evaluate the accuracy of ASGARD in single drug repurposing, we compared ASGARD to other repurposing methods that utilize bulk cell RNA-Seq (CLUE and DrInsight) or single-cell RNA-Seq data (Alakwaa's and Guo's) on a variety of diseases, including breast cancer, leukemia, and COVID-19. ASGARD performs much better than all these methods in predicting drugs/compounds (Figure 2, 3, Supplementary Figure 2). The performance of ASGARD is also robust across different sizes and proportions of cell populations, as well as differential expression levels (Supplementary Figure 3). Moreover, we highlight that ASGARD defines a novel drug score to summarize drug efficiency across multiple selected cell clusters. These important functions are missing in other simpler single-cell RNA-Seq drug reposition pipelines by Alakwaa and Guo. Both Alakwaa' and Luo's pipelines use the CLUE platform, a cloud-based platform developed by the LINCS Center for signature-gene based drug ranking[61]. Additionally, Luo's method uses log fold change as the additional threshold to filter the gene query. ASGARD on each cluster is related to DrInsight, a concordantly expressed genes (CEG) based, enhanced drug repurposing method compared to other signature-based searching methods[30]. It uses order statistics to directly measure the concordance (e.g. inverse association) between the disease data and drug-perturbed data and identifies concordantly expressed genes (CEGs). CEGS are used as features to further formulate an outlier sum statistic for drug selection, rather than the connectivity score (usually -90) based cut-off for drug selection. The CEG and outlier sum statistics contribute to higher performance in ASGARD.

ASGARD achieves drug ranking for the disease/patient by a novel drug score that evaluates the treatment efficacy across the user-selected cell clusters (Formula 1 in **Methods**). The prediction using the multi-cluster drug score shows a significantly (P-value < 0.05, student's t-test) better AUC than the prediction based on individual clusters (Figure 2, 3). It suggests that targeting an individual cell cluster isn't sufficient for successful drug prediction. Instead, targeting multiple essential diseased cell clusters is a more appropriate strategy for drug prediction. On the other hand, it is not ideal to propose drug repurposing using bulk RNA-seq, a mixture of all cells, as done by traditional methods (e.g. CLUE and DrInsight). Significant heterogeneity exists in different T cell populations; not all these cells play equal roles in the diseases [62,63], reflected by different gene expression responses to drug treatment[64]. ASGARD can distinguish more important T cell types from others and repurpose drugs accordingly, explaining why ASGARD has significantly (P-value <0.05, student's t-test) better AUC performance than traditional bulk methods (Figure 2B). Moreover, ASGARD also demonstrates variations in drug scores across different patients (Figure 4B, 5D, and 6D). This result stresses that personalized therapy is necessary for the best therapeutic effect, and utilizing single-cell sequencing information may help to achieve that.

Sparsity and heterogeneity are two major challenges in analyzing single-cell data, and usually cause false discoveries of differentially expressed genes[65]. Previous benchmark study showed that Seurat[18], DESeq2[27], edgeR[28], and Limma[26] are among the top methods in discovering differentially expressed genes using single-cell data[37]. We here compared the effect of these methods and different parameterization on the downstream drug repurposing, using AUC metric. AUC performance still varies with methods for single-cell differential expression (Figure 2A, Limma (Bayes) method showed the best average AUC performance compared to all other three methods. Within limma method, approaches that model mean-variance with the empirical Bayes approach (limma Bayes and limma trend) showed better AUCs than that with the precision weights approach (limma voom). Similar observations were observed in some of the comparisons in a benchmark study of single-cell differential expression[37]. The empirical Bayes approach is usually more powerful than the precision weights approach when the library sizes are not quite variable between samples[66]. Seurat, a method widely used in single-cell studies, has the 2nd best AUCs in general. In particular, the default mode of Wilcoxon rank-sum test in Seurat has slightly better average AUC than the t-test and logistic regression (LR) modes. Our comparison revealed that DE methods should be carefully selected according to the status of the dataset to achieve the best performance. Accordingly, ASGARD was designed as a flexible framework supporting various methods for single-cell differential gene expression analysis.

We chose breast cancer or leukemia datasets to illustrate the utilities of ASGARD, given the relative abundance of prior drug knowledge. FDA has approved many drugs predicted by ASGARD, such as neratinib and vinblastine for treating breast cancer [67,68] (Figure 4B, Supplementary Figure 4C), and tretinoin for treating leukemia [51] (Figure 5D). Vinblastine and neratinib were predicted for breast cancer in both TNBC patient and PDX datasets. Vinblastine is a vinca alkaloid that has been used in the treatment of metastatic breast cancer since the early 1980s. The regimen of vinblastine/mitomycin is an effective salvage regimen and an excellent first-line chemotherapeutic treatment for women with metastatic breast cancer[42]. Neratinib is a protein kinase inhibitor that was approved in July 2017 as an extended adjuvant therapy in breast cancer [68]. Recently, a randomized phase III clinical trial of 621 patients from 28 countries showed neratinib significantly improved the progression-free survival of patients with advanced breast cancer [69]. Tretinoin, also known as all-trans-retinoic acid (ATRA), is the first candidate predicted by ASGARD for leukemia. Tretinoin targets all significant pathways, such as p53 signaling, cell cycle, and apoptosis pathways, for each diseased cell cluster in leukemia patients (Figure 5E). These pathways play important roles in the survival of leukemia patients [54]. Consistent with our prediction, tretinoin was approved by the FDA to induce remission in patients with acute leukemia [51]. Tretinoin significantly improves the survival of acute leukemia[70]. Tretinoin with chemotherapy has become the standard treatment for acute leukemia, resulting in cure rates exceeding 80%[71]. The successful prediction of FDA-approved drugs supports the reliability of ASGARD.

Beyond the above described FDA-approved cases, ASGARD also predicts novel candidate drugs for breast cancer and leukemia. Crizotinib is a candidate from both TNBC patient and PDX model data (Figure 4B, Supplementary Figure 4C). Crizotinib is a receptor tyrosine kinase inhibitor that inhibits the growth, migration, and invasion of breast cancer cells in preclinical studies [44,45]. A case report showed the TNBC patient harboring the ALK fusion mutation had a dramatic response to crizotinib treatment [69]. It is also in a phase 2 clinical trial for treating patients with TNBC (ClinicalTrials.gov Identifier: NCT03620643). For leukemia, ASGARD predicts Vorinostat, a histone deacetylase (HDAC) inhibitor, as one of the novel candidate drugs. Vorinostat was approved by FDR for treating patients with progressive, persistent, or recurrent cutaneous T- cell lymphoma [72]. It induces cell apoptosis in one T cell leukemia cell line in vitro [73], and improves the outcome of acute T cell lymphoblastic leukemia in animal models [74]. The results of a recent clinical trial (ClinicalTrials.gov Identifier: NCT04467931) show vorinostat is a promising candidate drug for T cell acute lymphoblastic leukemia [75].

Since ASGARD repurposes candidate drugs to reverse "diseased" cells to "normal" cells, it's important to set proper controls according to the aim of the study. The best controls are arguably from the normal tissues of the same patient, and the next best ones are from the patients without such a disease but matched on other major confounders. Although consortiums such as Human Cell Atlas aim to obtain normal tissues from clinically healthy samples, it may not be easy to obtain normal samples for some diseases. Under such scenarios, samples from the very early stage of the disease or controls from tissues with the most relevant origin could be substitutes, until the data from the normal tissues are available. Additionally, a recent report has proposed using deep-learning based approach to identify the best reference control tissue, an attractive strategy that relies much less on prior-assumptions [76]. Additionally, ASGARD built the drug reference using drug responses data from LINCS L1000 project[29], which were collected from 98 cell lines. We divided the drug reference into several tissue specific drug references according to the tissue origin of the cell lines. It's highly recommended to select tissue specific drug references when using ASGARD. For example, for drug repurposing in breast cancer, it is best to use drug responses collected from breast cell lines. If there isn't a proper cell-type specific reference for the target disease, it might be worthy identifying the cell line whose base-line gene expression profiles are most similar to "control" samples, after adjusting for systematic differences between cell line vs. primary tissues (eg. using transfer learning).

Altogether, this study shows clear evidence that ASGARD defines a single-cell-based reliable drug score for repurposing confident drugs, which were approved or in clinical trials for breast cancer, leukemia, and COVID-19, respectively. It also provides new applications for drugs that warrant further clinical studies. In all, ASGARD is a single-cell guided pipeline with significant potential to recommend repurposeful drugs.

**Funding**

This research was supported by the National Institute of Environmental Health Sciences through funds provided by the trans-NIH Big Data to Knowledge (BD2K) initiative [K01ES025434]; the US National Library of Medicine [R01 LM012373, R01 LM12907]; and the National Institute of Child Health and Human Development [R01 HD084633; to L.X. Garmire].

**Competing interests**

The authors declare that they have no competing interests.

**Data and materials availability**

ScRNA-Seq data are available in Gene Expression Omnibus (Accession number: GSE161529, GSE123926, GSE113197, GSE132509, GSE158055, and GSE145926). Phase I LINCS L1000 data are available in Gene Expression Omnibus (Accession number: GSE92742). Phase II LINCS L1000 data are available in Gene Expression Omnibus (Accession number GSE70138). ASGARD is available as an R package on GitHub (https://github.com/lanagarmire/ASGARD). Scripts used in this study are available on GitHub (https://github.com/lanagarmire/Single-cell-drug-repositioning).

**Figure Legends**

Figure 1

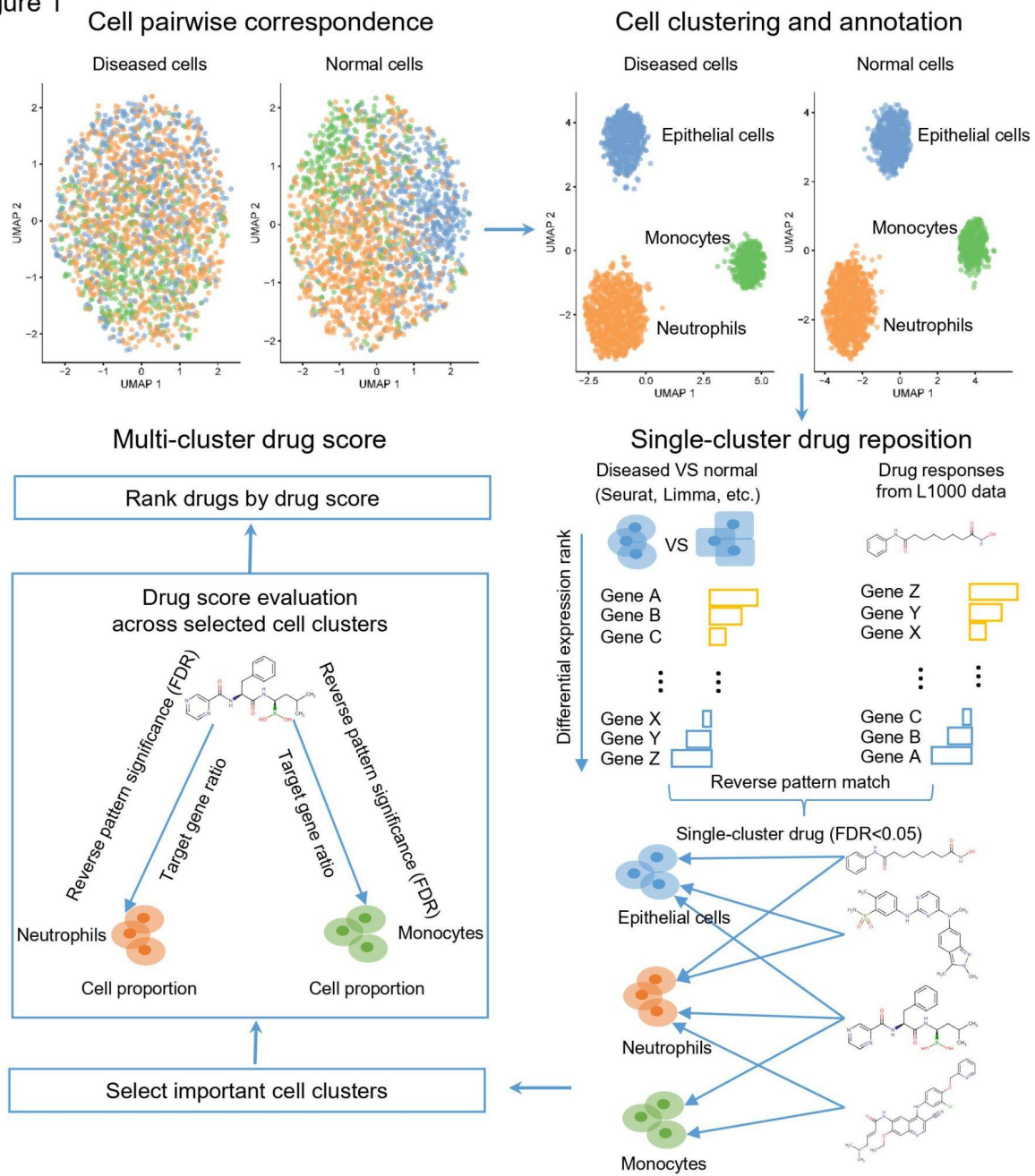

**Figure 1. The workflow of the ASGARD drug repurposing pipeline.** The workflow of the ASGARD pipeline. Diseased and normal cells are paired according to "anchor" genes that are expressed consistently between the two types of cells. The differentially expressed (DE) genes

are identified between diseased and normal cells, either within a cluster or within a cell type.
Using the consistent DE genes as the input, potential drugs that significantly reverse the pattern
of DE genes are identified, using the Kolmogorov–Smirnov (K-S) test with Benjamini-Hochberg
(BH) false discovery rate (FDR) adjustment. Next ASGARD estimates and ranks the drug
scores for single drugs, by targeting specific cell cluster(s) or all cell clusters.

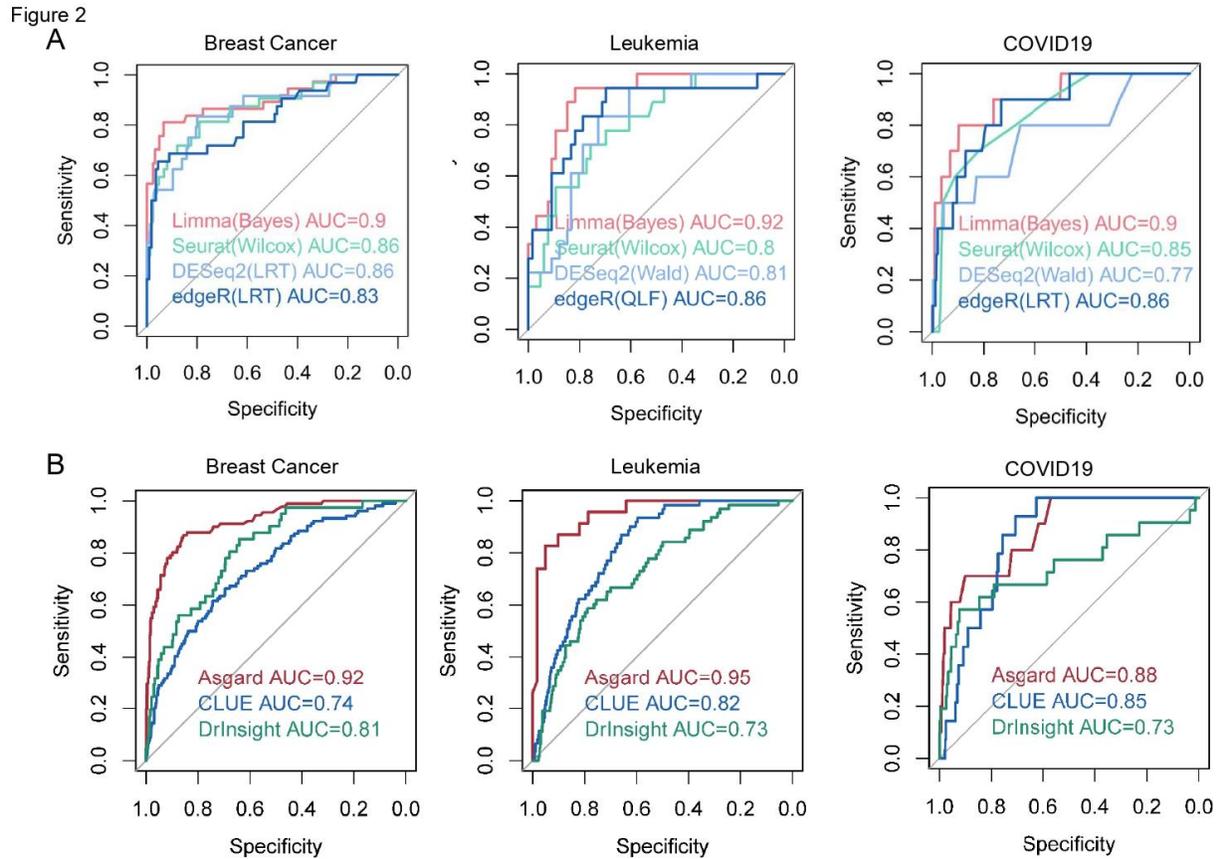

**Figure 2. Comparing ASGARD to bulk-cell-based repurposing methods.** The receiver operating characteristic (ROC) curves and area under curve (AUC) scores of the ASGARD, on advanced metastatic breast cancer, acute lymphoblastic leukemia, and coronavirus disease 2019 (COVID-19), respectively. (**A**) Comparison among different DE analysis methods Limma, DESeq2, Seurat, and edgeR using the best-performing mode in each. (**B**) Comparison of ASGARD and bulk-sample based drug repurposing methods (CLUE and DrInsight), using the same three diseases as in **(A)**. The single-cell RNA-Seq data were aggregated to pseudo-bulk

RNA-Seq data as the input of the bulk-sample based methods.

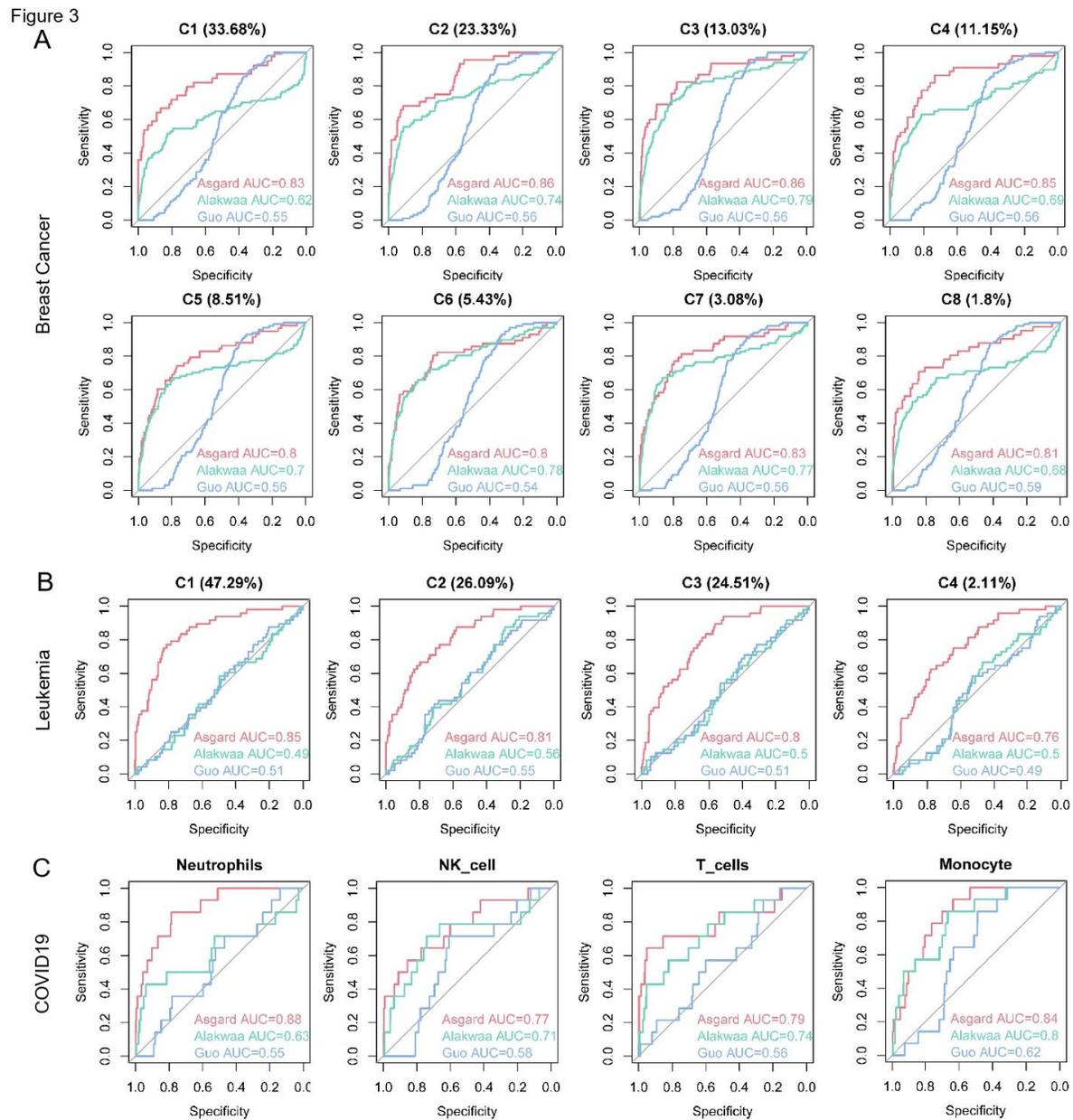

**Figure 3. Comparing ASGARD to other single-cell-based repurposing methods.**

ROC curves and AUC scores of the ASGARD and other published pipelines (Alakwaa's pipeline and Guo's pipeline). The results of drug/compound repurposing are shown on every cell cluster of the metastatic breast cancer dataset **(A)**, every cell cluster of the acute lymphoblastic leukemia dataset **(B)** and 4 clusters with increased cell proportions in the decreased severe vs cured severe COVID-19 patients **(C)**. The proportion of each single-cell cluster is shown in the brackets above each plot.

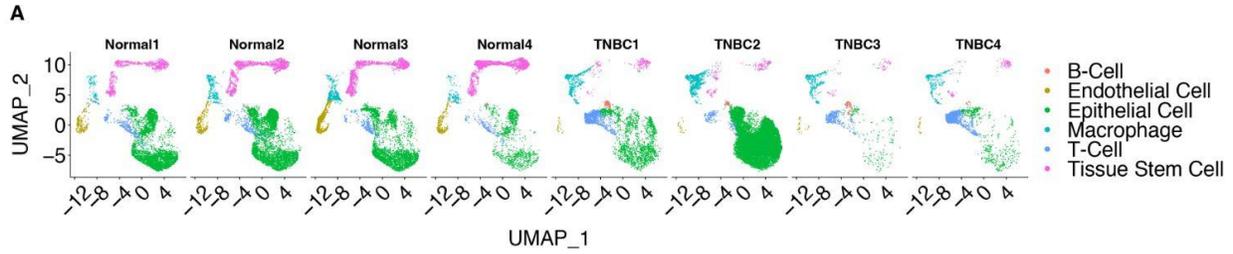
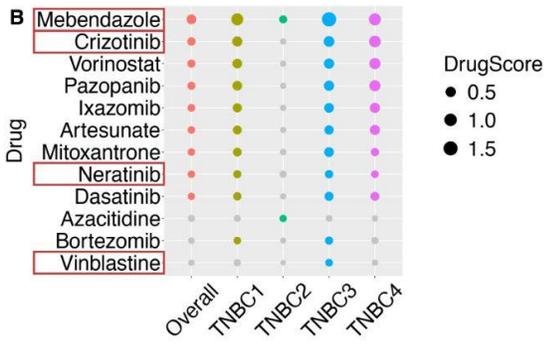
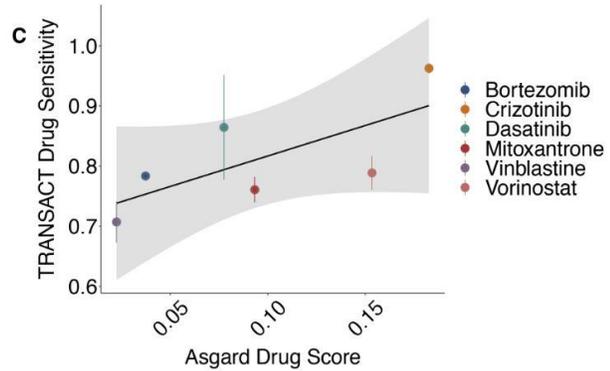
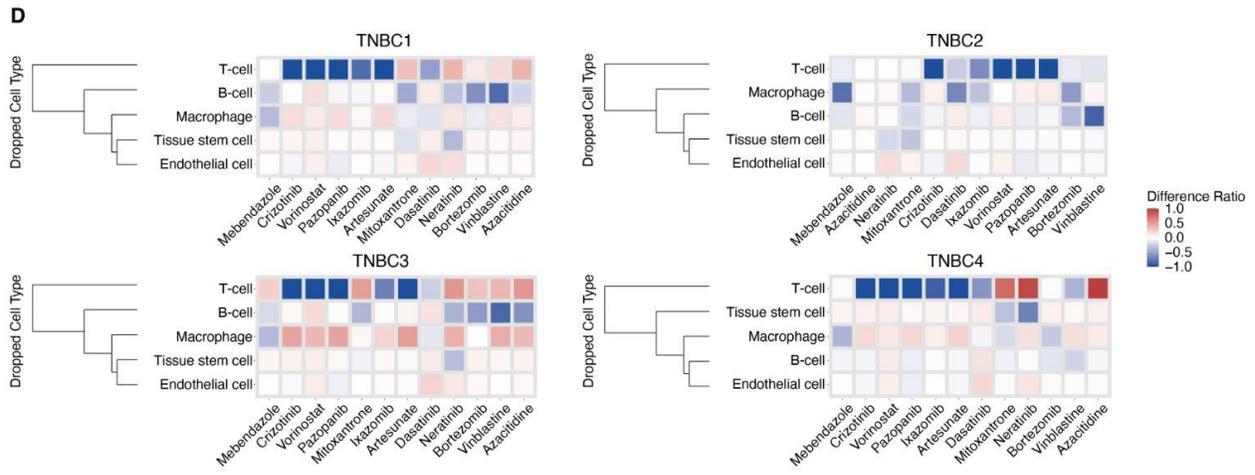
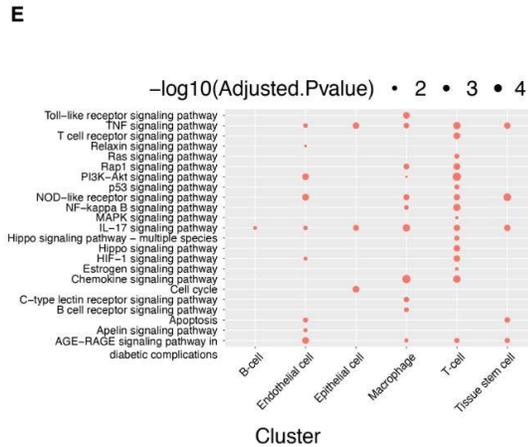
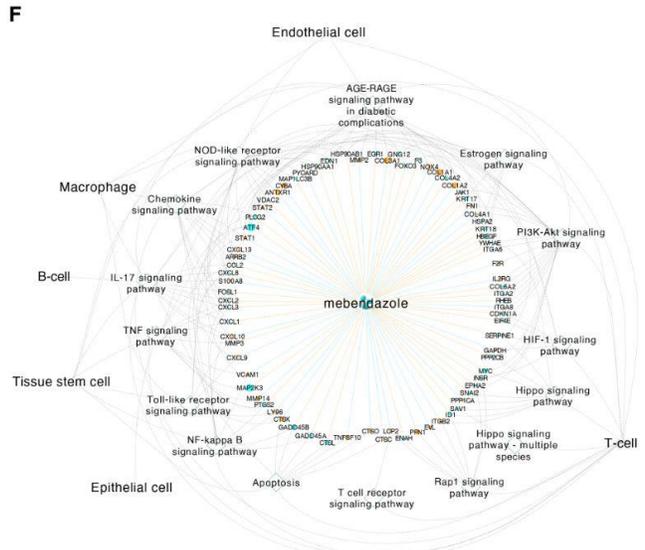

**Figure 4. Drug repurposing in TNBC patient samples. (A)** UMAP plots of single-cell data from four TNBC patient samples and four control samples from the same study. **(B)** The overall drug scores combining all TNBC samples as well as drug scores in each TNBC sample, among the top-ranked significant single drugs (FDR<0.05). Gray color indicates a lack of significance for a particular drug in an individual. The four drugs in the red boxes overlapped with the top drugs predicted by ASGARD using PDX models of TNBC patients. **(C)** Comparison between TRANSACT drug sensitivity and ASGARD drug score. **(D)** Heatmap of the drug score changes in the cell-type-specific drop-one-out experiment, where each cell type in the tumor microenvironment is removed at a time. **(E)** Pathway enrichment analysis (TNBC vs. normal) for each cell cluster **(F)** The top drug candidate Mebendazole, its target genes, pathways, and single-cell clusters. Orange node: up-regulated gene (logFC>1 and adjusted P-value<0.05) . Blue node: down-regulated gene (logFC<-1 and adjusted P-value<0.05). Orange solid edge: drug stimulates gene expression. Blue solid edge: drug inhibits gene expression. The width of the edge is proportional to the strength of the drug effect. Gray dotted edge: gene belonging to a pathway. Gray backward slash: pathway significant in a cell cluster.

Figure 5

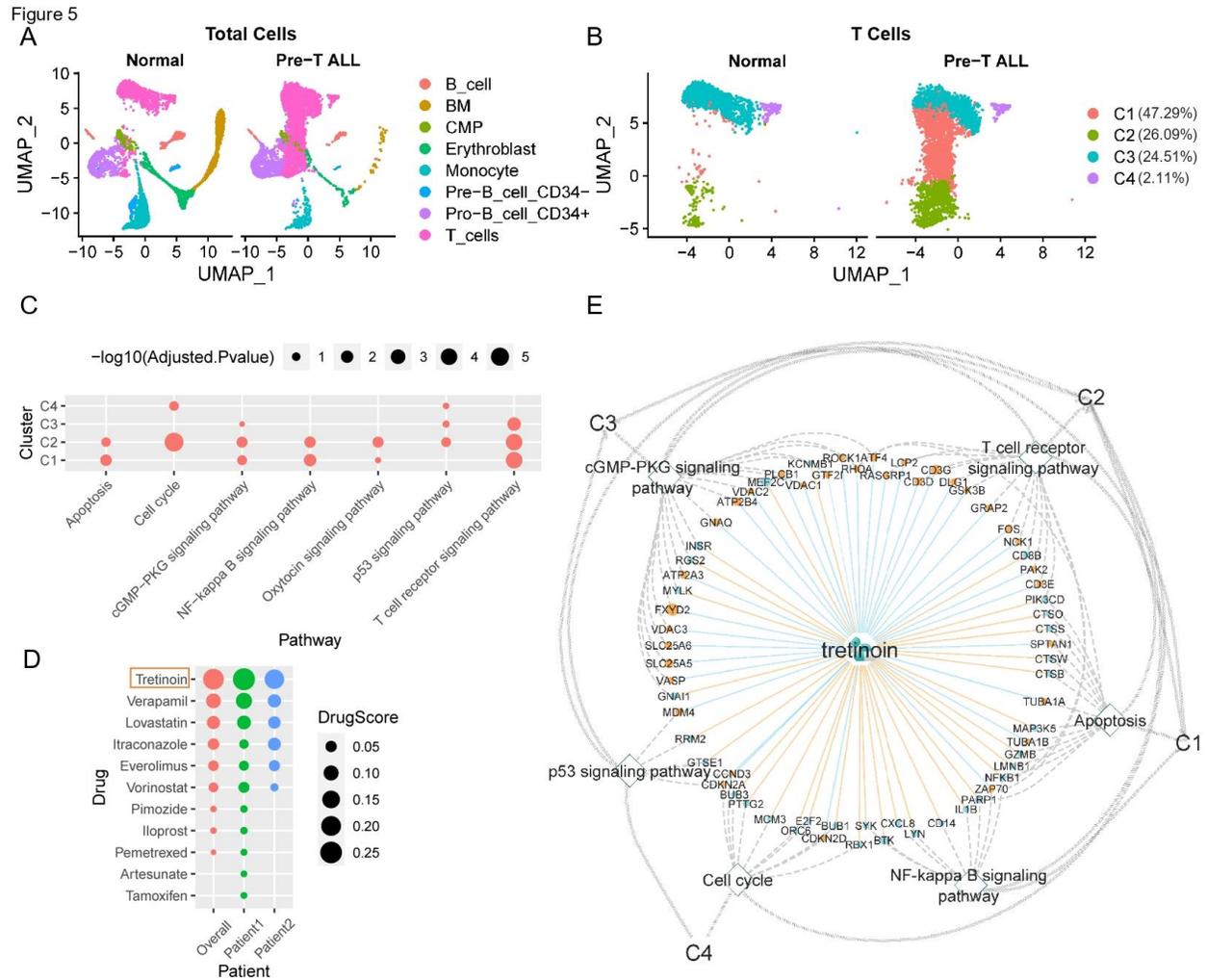

**Figure 5. Drug repurposing for precursor T cell acute lymphoblastic leukemia (Pre-T ALL). (A)** UMAP plots of all cells from 3 normal controls and 2 Pre-T ALL samples. **(B)** UMAP plots of T cell clusters from normal controls and Pre-T ALL samples. **(C)** Pathway enrichment analysis (leukemia vs normal) for each T cell cluster. **(D)** The overall drug score and drug score in each Pre-T ALL patient, among top-ranked significant drugs (FDR<0.05). Drug approved for leukemia treatment by the FDA is tretinoin. **(E)** The drug candidate tretinoin, its target genes, pathways, and single-cell clusters. All labels and their annotations are the same as Figure 4F.

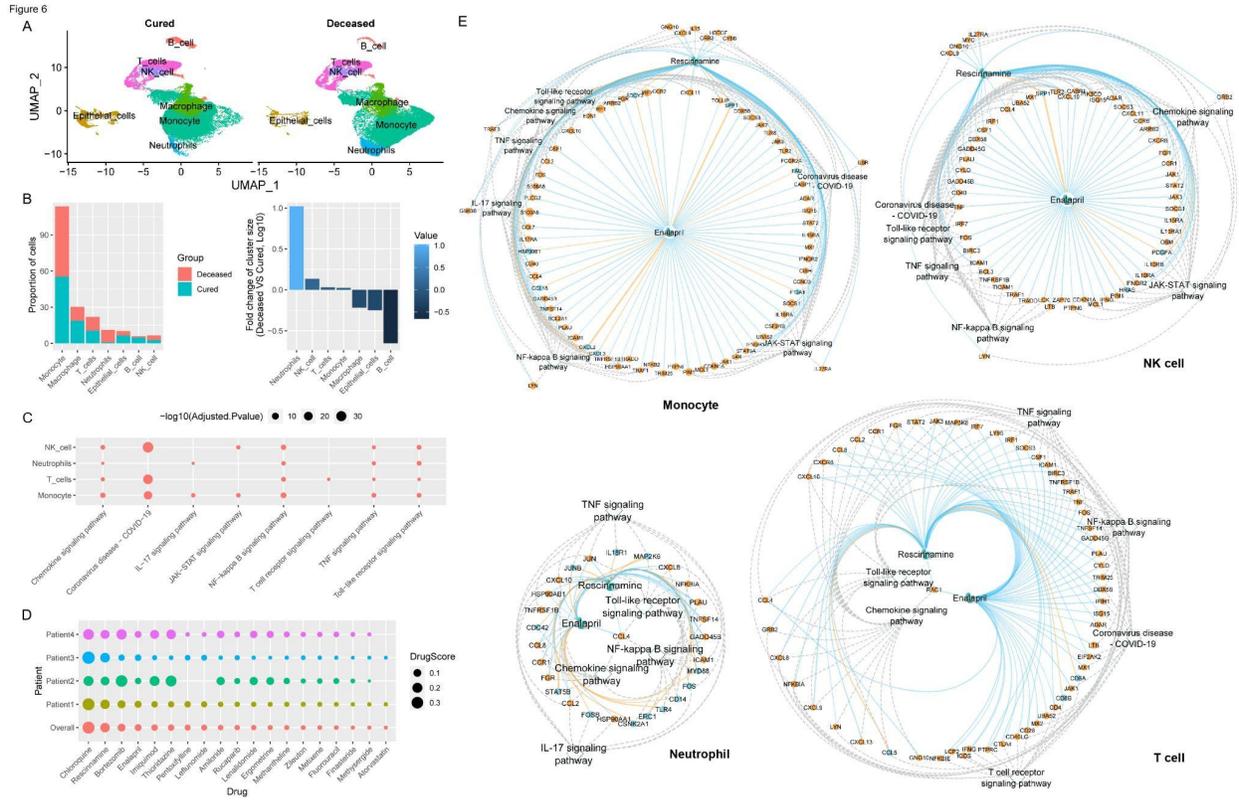

**Figure 6. Drug repurposing for reducing mortality of severe COVID-19 patients. (A)** Single-cell populations of bronchoalveolar immune cells in 11 cured and four deceased severe COVID-19 patients, respectively. **(B)** The proportions of cell type in (left) and log10 transformed fold changes in deceased over the cured state (right) of the single-cell populations in (A). **(C)** Pathway enrichment analysis (deceased severe vs. cured severe) for neutrophil, NK cell, T cell, and monocyte, respectively. **(D)** The drug scores in the four deceased severe COVID-19 patients and all four patients among top-ranked significant drugs (FDR<0.05). **(E)** The drug candidate rescinnamine and enalapril, their target genes and pathways in the monocyte, NK cell, T cell, and neutrophil, respectively, from severe COVID-19 patients. All labels and their annotations are the same as in Figure 4F.

**Supplementary Materials**

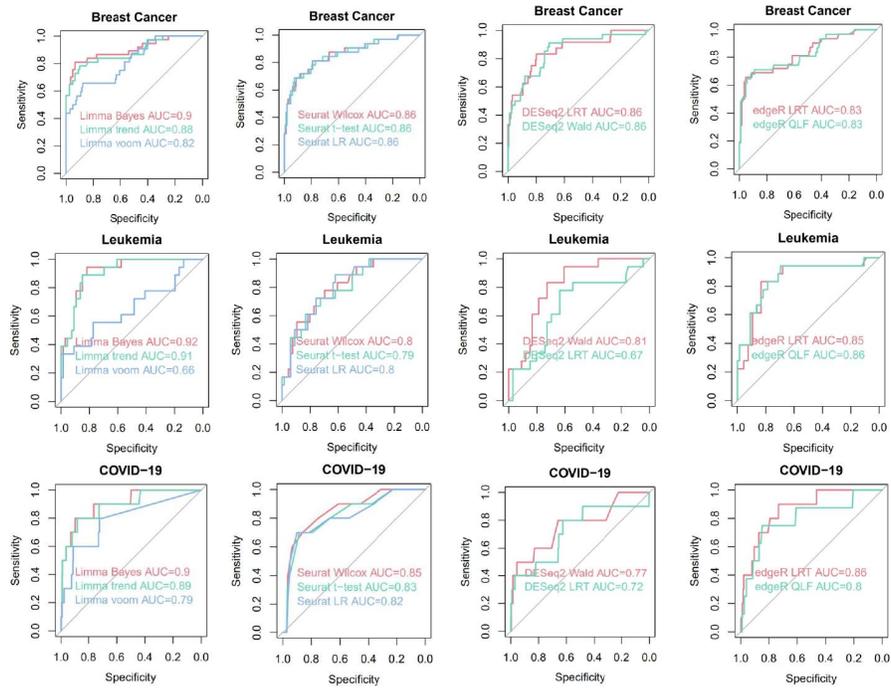

**Supplementary Figure 1. The impact of single-cell differential expression methods on ASGARD performances.** The receiver operating characteristic (ROC) curves and area under curve (AUC) scores of the ASGARD, using DE analysis methods (Limma, DESeq2, Seurat, and edgeR) with tuned parameters. The tests are done on advanced metastatic breast cancer, acute lymphoblastic leukemia, and coronavirus disease 2019 (COVID-19), respectively.

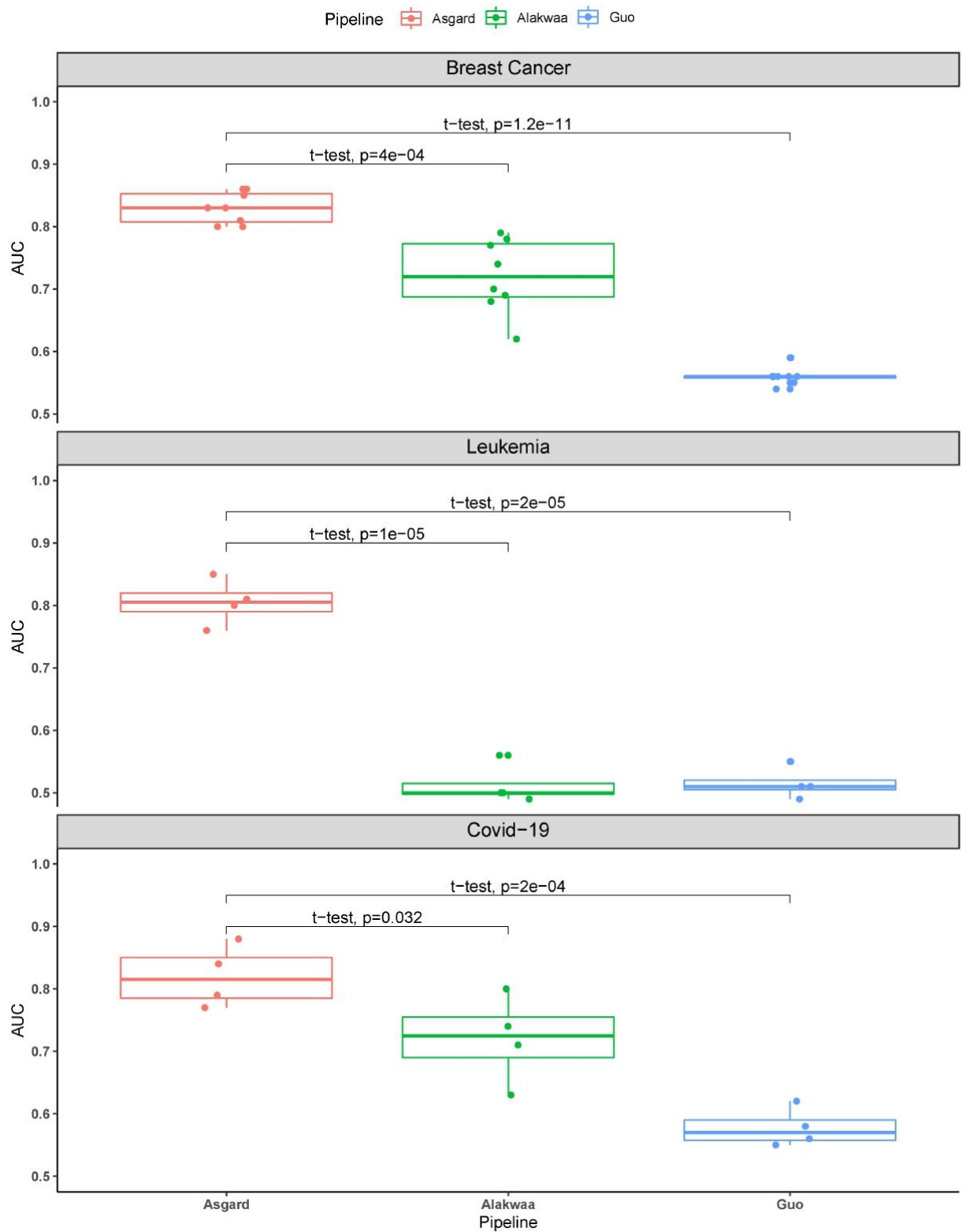

**Supplementary Figure 2.** Boxplots on AUCs of the three single-cell-based repurposing pipelines, based on all cell clusters, as shown in Figure 3.

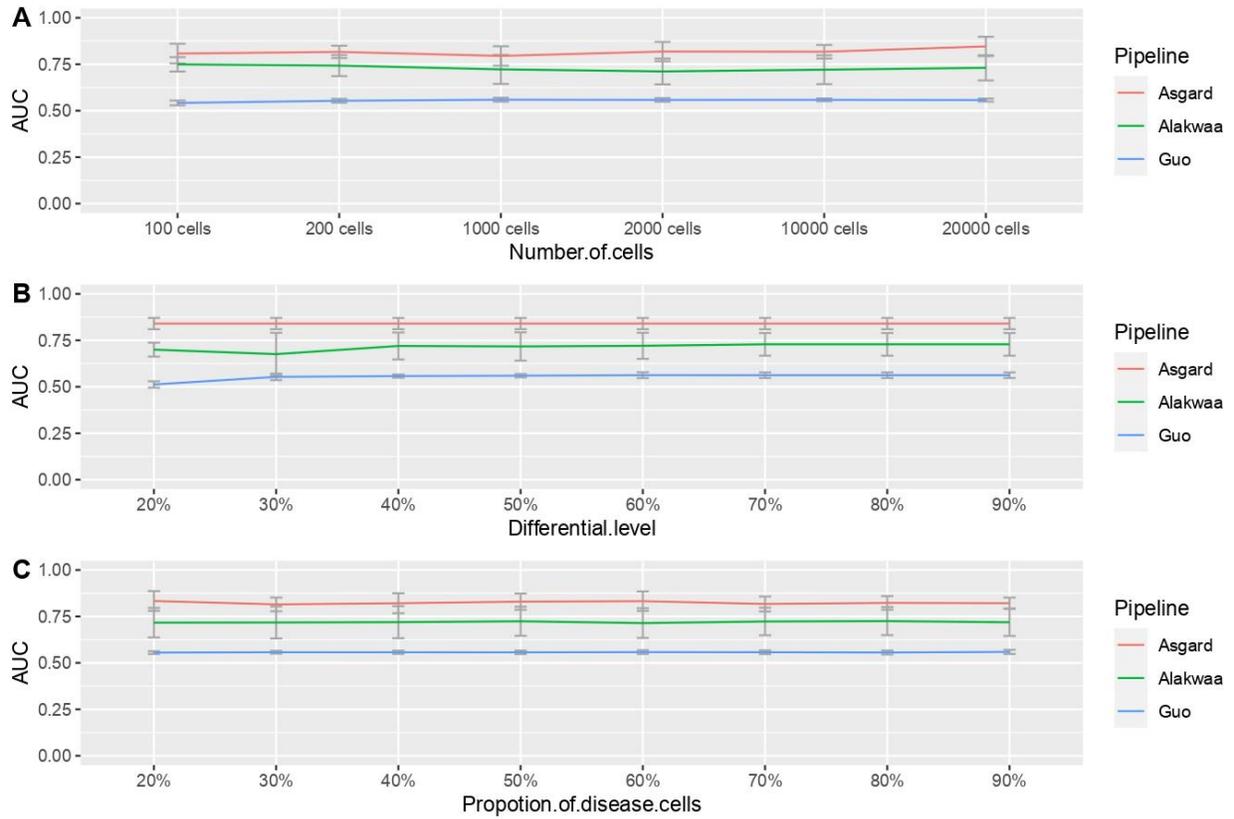

**Supplementary Figure 3.** AUC scores of the three single-cell-based repurposing pipelines, using simulation data adapted from the real datasets. **(A)** The effect of varying total cell sizes from 100 cells to 20000 cells. **(B)** The effect of varying differential expression levels from 20% to 90%. **(C)** The effect of varying the proportions of diseased cells ranges from 20% to 90%.

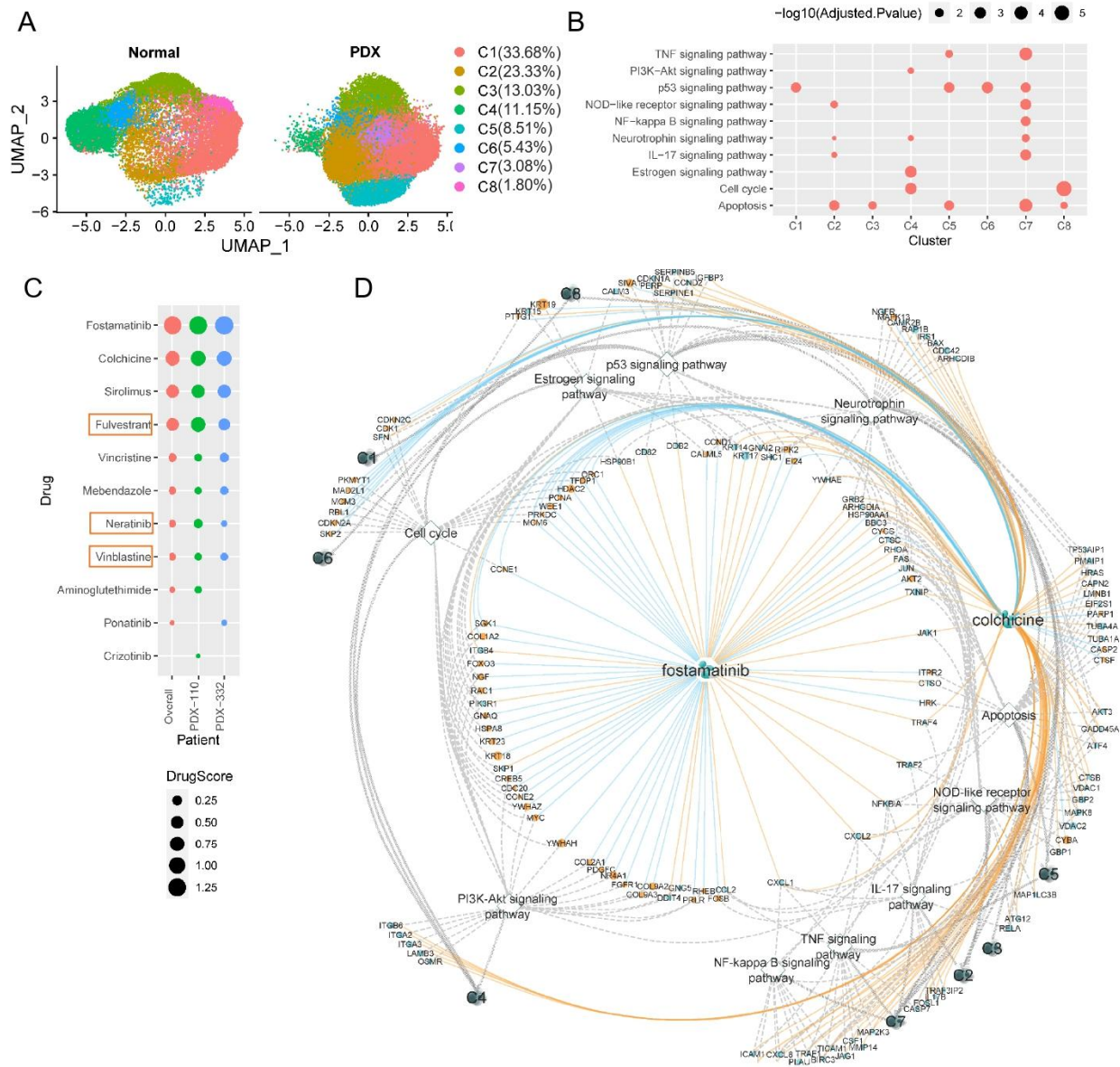

**Supplementary Figure 4. Drug repurposing in Patient-Derived Xenograft (PDX) models derived from advanced metastatic TNBC patients. (A)** UMAP plots of single-cell data from 3 normal controls and 2 breast cancer PDX samples. **(B)** Pathway enrichment analysis (breast cancer vs normal) for each single-cell cluster. **(C)** The overall drug score combining both PDX models and drug score in each breast cancer PDX model, among top-ranked significant single drugs (FDR<0.05). Drugs approved for breast cancer treatment by the FDA are labeled in red

boxes. **(D)** The drug candidates fostamatinib and colchicine, their target genes, pathways, and single-cell clusters. All labels and their annotations are the same as Figure 4F.

**Supplementary Note 1.** Analysis of PDX breast cancer model.

We collected scRNA-seq data from 24,741 epithelial cells of advanced metastatic breast cancer Patient-Derived Xenografts (PDXs) models [11] and 16,998 epithelial cells from normal breast tissues [21]. After preprocessing, all cancer cells and 16,954 normal cells were paired and clustered into 8 populations (Supplementary Figure 3A). Cluster 1 (C1) is the largest one covering 33.68% of cells, while cluster 8 (C8) is the smallest one accounting for only 1.8% of cells (Supplementary Figure 3A). The differentially expressed genes (adjusted P-value <0.05, cancer vs normal) in the clusters are significantly enriched in 10 well-known breast cancer-related pathways, including apoptosis, cell cycle, estrogen signaling, IL−17 signaling, neurotrophin signaling, NF−kappa B signaling, NOD−like receptor signaling, p53 signaling, PI3K−Akt signaling and TNF signaling pathways (Supplementary Figure 3B). Cluster 7 (C7) has the largest number of 7 significant pathways, while C1 and C6 each have only 1 significant pathway.

We first applied ASGARD for multi-cluster drug repurposing prediction and predicted 11 drugs (FDR<0.05 and overall drug score >0.99 quantiles) for advanced metastatic breast cancer (Supplementary Figure 3C, Supplementary Table 4). Fostamatinib is the top 1 drug candidate (Supplementary Figure 3C). It is a tyrosine kinase inhibitor medication approved for the treatment of chronic immune thrombocytopenia [77]. Colchicine, the second best candidate, is an alkaloid approved for treating the inflammatory symptoms of familial Mediterranean fever [78]. Both fostamatinib and colchicine have shown antitumor and anti-metastasis effects in animal models of breast cancer [79,80]. Moreover, the 4th candidate fulvestrant and 7th candidate neratinib have been approved by the Food and Drug Administration (FDA) for breast cancer treatment [58,59].

To explore the potential molecular mechanisms of the top 2 candidates, we next investigated the target genes and pathways of fostamatinib and colchicine across the eight cell clusters (Supplementary Figure 3D). Fostamatinib and colchicine both target all the significant pathways in each cluster. Fostamatinib and colchicine are complementary in targeting genes of these pathways. Among the 143 target genes from these significant pathways, only 29 target genes are shared by fostamatinib and colchicine (Supplementary Figure 3D). The fostamatinib and colchicine also show biologically synergistic targeting of multiple genes on the same significant pathways. For example, fostamatinib inhibits Cyclin D1 (CCND1) to produce G1 arrest in the p53 signaling pathway, while colchicine inhibits Cyclin-dependent kinase 1 (CDK1) to produce G2 arrest in the p53 signaling pathway and cell cycle pathway [81] (Supplementary Figure 3D). Additionally, the drug scores of top drug candidates vary from one PDX model to another (Supplementary Figure 3D), demonstrating that ASGARD is a forward-looking precision medicine strategy *in silico*.

**Supplementary Table 1.** FDA-approved drugs and compounds used in advanced clinical trials or have been proven effective in animal models

**Supplementary Table 2**. Predicted drugs for leukemia.

**Supplementary Table 3**. Predicted drugs for COVID-19.

**Supplementary Table 4**. Predicted drugs for breast cancer PDX model.

195–200 (1997).

71. Lo-Coco, F. *et al.* Retinoic acid and arsenic trioxide for acute promyelocytic leukemia. *N. Engl. J. Med.* **369**, 111–121 (2013).

72. Vorinostat. in *LiverTox: Clinical and Research Information on Drug-Induced Liver Injury* (National Institute of Diabetes and Digestive and Kidney Diseases, 2020).

73. Gao, M. *et al.* Therapeutic potential and functional interaction of carfilzomib and vorinostat in T-cell leukemia/lymphoma. *Oncotarget* **7**, 29102–29115 (2016).

74. Jing, B. *et al.* Vorinostat and quinacrine have synergistic effects in T-cell acute lymphoblastic leukemia through reactive oxygen species increase and mitophagy inhibition. *Cell Death Dis.* **9**, 589 (2018).

75. Siddiqi, T. *et al.* Phase 1 study of the Aurora kinase A inhibitor alisertib (MLN8237) combined with the histone deacetylase inhibitor vorinostat in lymphoid malignancies. *Leuk. Lymphoma* **61**, 309–317 (2020).

76. Zeng, B. *et al.* OCTAD: an open workspace for virtually screening therapeutics targeting precise cancer patient groups using gene expression features. *Nat. Protoc.* **16**, 728–753 (2021).